\documentclass[reprint, aps, superscriptaddress, prab]{revtex4-2}
\usepackage[utf8]{inputenc}
\usepackage{graphicx}
\usepackage{amsmath, amssymb}
\usepackage{xcolor}
\usepackage{upgreek}
\usepackage{hyperref}
\usepackage{siunitx}

\begin{document}

    \title{Influence of Laguerre-Gaussian pulse mode purity on positron acceleration in nonlinear wakefields}

	\author{Lars Reichwein}
	\email{l.reichwein@fz-juelich.de}
    \affiliation{Peter Gr\"{u}nberg Institut (PGI-6), Forschungszentrum J\"{u}lich, 52425 J\"{u}lich, Germany}
    
	\date{\today}
	
	\begin{abstract}
    The acceleration of positrons in wakefields remains a crucial challenge. One proposal has been to use Laguerre-Gaussian laser pulses as drivers, which provide corresponding accelerating and focusing regions due to their inherent field structure. Here, we study the effect of mode order and mode purity on the acceleration process by means of particle-in-cell simulations. We find that some mode impurities can be mitigated during the propagation of the wake through the plasma, but also that phase difference between the modes that the total pulse is composed of can strongly affect the acceleration process.
	\end{abstract}

\maketitle

\section{Introduction}

Wakefield acceleration is a promising alternative to conventional accelerators, as wakefields can provide significantly stronger accelerating gradients \cite{Esarey2009}. To drive a wakefield, either a high-intensity, short laser pulse (LWFA) or an energetic particle beam (PWFA) can be utilized. 
In the laser-driven case, the ponderomotive force pushes out electrons, leaving behind a cavity (``bubble'') for a sufficiently strong and short laser pulse \cite{Pukhov2002}. This bubble structure follows the driving pulse, and exhibits linear accelerating and focusing fields. 
Electrons can be injected into the wake by a variety of mechanisms, e.g. self-injection \cite{Mangles2012}, density ramps \cite{Ekerfelt2017} or ionization injection \cite{Chen2006}.
Over the years, various schemes have produced high-energy \cite{Picksley2024} or ultra-low-emittance electron beams \cite{Hidding2012}.

For positrons, however, the situation is different: the wakefield regions that are accelerating and focusing for electrons, are decelerating and defocusing for positrons. 
This challenge has led to the proposal of various schemes which rely, e.g., on different driver shapes \cite{Jain2015, Vieira2014}, specific channel structures \cite{Diederichs2019, Reichwein2022} or going to different acceleration methods altogether, such as direct laser acceleration  \cite{Martinez2025}. An overview of the state of the art for wakefield-based positron acceleration is found in Ref. \cite{Cao2024}.

One particularly interesting proposal by Vieira and Mendon{\c{c}}a for positron acceleration in wakefields utilizes Laguerre-Gaussian (LG) laser pulses to accelerate positrons efficiently \cite{Vieira2014}.
Laguerre-Gaussian pulses can carry orbital angular momentum (OAM) and are characterized by their azimuthal index $\ell$ and their radial index $p$, in short $(\ell, p)$. 
Due to their field structure, some of these modes are referred to as ``donut modes''.
In Ref. \cite{Vieira2014}, this structure of a (1,0) mode is utilized to form a wakefield region which focuses and accelerates positrons on-axis.

Since then, more work on LG pulses for particle acceleration has been conducted. Besides positron acceleration, LG pulses have also been proposed for ion acceleration \cite{Zhang2014, Hu2022, Willim2023, Reichwein2025}, as they are able to produce more collimated ion beams than comparable Gaussian pulses. The changes in induced field structure (in particular the azimuthal magnetic field) also enable the acceleration of spin-polarized electrons and ions \cite{Wu2019lwfa, Reichwein2026}.

While the applications of Laguerre-Gaussian laser pulses in plasma-based accelerators are plentiful, experimental realization of high-intensity and high-quality pulses remains challenging.
Currently, intensities in the range of $6.3\times 10^{19} \, \si{\watt\per\square\centi\meter}$ have been demonstrated \cite{Wang2019}. 
Some theoretical proposals on how to generate higher-power pulses exist, like plasma volume holograms \cite{Lehmann2019} or stimulated Raman amplification \cite{Vieira2016}.

Provided that sufficiently high-intensity LG pulses can be generated, the other crucial concern for many of the acceleration schemes relying on the LG geometry is mode purity: with commonly utilized spiral phase plates a superposition of several modes is expected rather than a pure, single $(\ell, p)$. This is already apparent in lower-power fiber lasers \cite{Zalesskaia2026} and likely to influence the laser-plasma interaction.

In this paper, we study the acceleration of positron beams using LG pulses of various mode purity based on the scheme of Ref. \cite{Vieira2014} using particle-in-cell simulations. We compare the acceleration of positrons using pure (1,0) and (1,1) modes, as well as a combination of the two in section \ref{sec:pic}. The simulations show that the final beam energy as well as its angular divergence are influenced by mode purity. The relative phase between the modes is identified as a crucial parameter for the apparent improvements.
Moreover, the simulations indicate the mode impurities can be partially mitigated during propagation through plasma.
In section \ref{sec:ana}, we derive the radial intensity profile of the mode combination and discuss the relevance of the phase difference for beam propagation.

\section{Particle-in-cell simulations} \label{sec:pic}

\begin{figure*}[htpb]
    \centering
    \includegraphics[width=\textwidth]{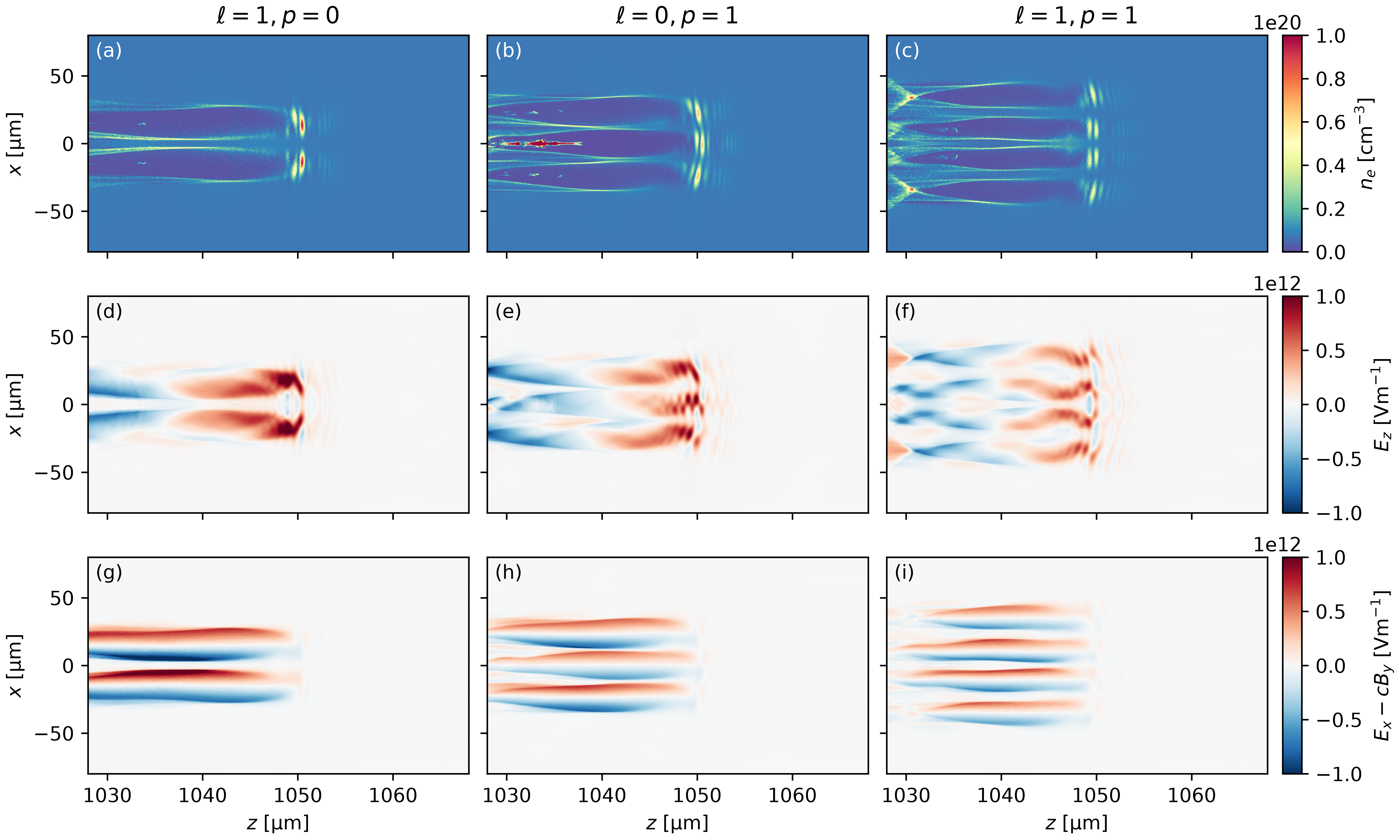}
    \caption{\label{fig:modes2d}Overview of electron density (a)-(c), longitudinal electric field $E_z$ (d)-(f), and focusing field $E_x - c B_y$ (g)-(i) for different, pure modes at the final time step of the simulation. Note that the colorbars are clipped for better visibility.}
\end{figure*}

We conduct particle-in-cell (PIC) simulations using the code \textsc{fbpic} \cite{fbpic} which uses a spectral cylindrical representation of the grid. The simulation domain has a size of 40 {\textmu}m $\times$ 80 {\textmu}m with a resolution of 1000 $\times$ 250 cells, and three azimuthal modes, to accurately simulate the propagation of the Laguerre-Gaussian modes.
Throughout the simulation, we keep the homogeneous background electron density fixed at $n_0 = 7.7 \times 10^{18} \si{\per\centi\meter\cubed}$.

The driving laser pulse has a normalized laser vector potential $a_0 = 10$ (linear polarization), a focal spot size of $w_0 = 12 \, \si{\micro\meter}$ and a pulse duration of $10 \, \si{\femto\second}$. The mode (composition) will be varied throughout different simulation runs, however in such a way, that the pulse energy remains fixed for all runs.

A cylindrical positron beam with $\gamma = 200$ (and initially zero transverse momentum) is placed 7.5 {\textmu}m behind the laser pulse. This beam has a length of 3.84 {\textmu}m and a width of 5.76 {\textmu}m, which is based on the parameters of Vieira \cite{Vieira2014}.

Upon entering the plasma region, the $(\ell, p) = (1,0) $ laser pulse forms a donut-shaped wakefield. The region around the optical axis $r=0$ provides a field structure capable of focusing and accelerating on-axis positrons.
For other modes, as well as different degrees of mode purity, the generated structure will be more complicated, as discussed below. We  first compare the influence of different, pure modes on the acceleration. Subsequently, we study the effects of mode impurities on the acceleration by superposing two different modes.

\subsection{Influence of mode order}

We compare three different LG modes, $(\ell, p) = (1,0)$, $(0,1)$ and $(1,1)$. We leave out the study of any higher-order ($\ell > 1$) modes here, as generating these at high intensities and purity would be extremely challenging.

The generated wakefield structure, as well as the induced accelerating field $E_z$ and the focusing force $E_x - c B_y$ is depicted in Fig. \ref{fig:modes2d}. In case of the $(1,0)$ mode, the usual donut wakefield can be seen.
The choice of mode order influences the number of wakefield lobes in the $x$-$z$ cross-section: for (0,1), the number of lobes increases to 3, while for the (1,1) mode four lobes are observed towards the end of the simulation (approx. 1 mm propagation through homogeneous plasma).

Both modes with $\ell = 1$ are capable of accelerating positrons over the full acceleration distance of 1 mm, whereas the mode $(0,1)$ ejects all positrons over time. This is due to the fact that the (0,1) mode produces a wakefield lobe around $r = 0$, where the positrons are initially located. Inside a single lobe, the electromagnetic fields correspond to that of a conventional (i.e. Gaussian-pulse-produced wakefield) meaning that they are defocused and cannot be trapped by the outer wakefield regions.

\begin{figure}[htpb]
    \centering
    \includegraphics[width=\columnwidth]{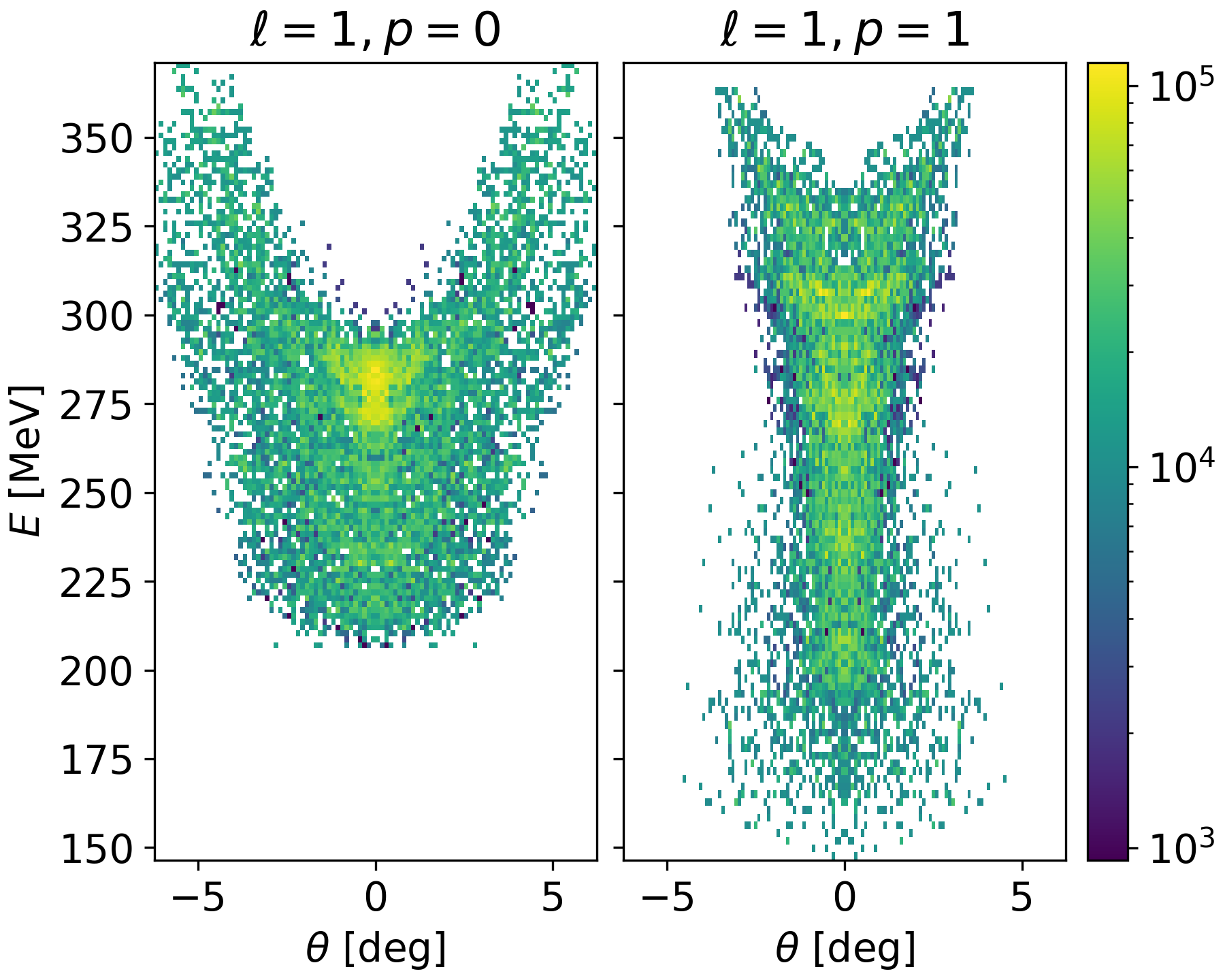}
    \caption{\label{fig:mode_order_phase} Angle-energy phase space for two different, pure modes. The case of $(\ell,p) = (0,1)$ is not plotted here as all positrons are lost.}
\end{figure}

The angle-energy phase space for the modes is shown in Fig. \ref{fig:mode_order_phase}. For both the (1,0) and the (1,1) mode, the maximum observed energy exceeds 350 MeV. However, for the (1,0) mode, the highest-energy positrons are located around $\theta = \pm 5^\circ$, while lower angular divergence is observed for (1,1). For the (1,0) mode, the bulk of the accelerated positrons sits around $\theta = 0^\circ$ with an energy of approx. 275 MeV.

The emittance of the positron beam is $\epsilon_x = 5.1 \times 10^{-5} \si{\meter\radian}, \; \epsilon_y = 2.9 \times 10^{-5} \si{\meter\radian}$ for (1,0) and $\epsilon_x = 1.6 \times 10^{-5} \si{\meter\radian}, \; \epsilon_y = 1.9 \times 10^{-5} \si{\meter\radian}$, i.e. the emittance is more asymmetric for (1,0), but between the two simulations, the emittance is within the same order of magnitude.

What has not been explicitly considered here, is the focusing of the different LG modes throughout their propagation: it is known (see, e.g., \cite{Willim2024}) that the threshold for self-focusing is different between LG modes and conventional Gaussian pulses. A future, separate study could investigate the optimization of LG pulse propagation through plasma channels instead of homogeneous plasma for stabilized positron acceleration. Instead, we focus on the effects of mode purity in the following.

\subsection{Influence of mode purity}

To study the influence of mode purity, we superpose two different modes $(1,0)$ and $(1,1)$. In the following, we use the variable $\eta$ to denote the degree of purity. Based on the normalization of \textsc{fbpic} for the donut-like LG modes, we give mode $(1,0)$ an $\tilde{a}_0 = \sqrt{\eta} a_0 $ and mode $(1,1)$ $\tilde{a}_0 = \sqrt{1 - \eta} a_0$ to ensure constant pulse energy. 

We compare the pure mode ($\eta = 1$) of the previous simulations with a run for $\eta \in \lbrace 0.3, 0.5, 0.7, 0.9 \rbrace$. 

\begin{figure*}[htpb]
    \centering
    \includegraphics[width=\textwidth]{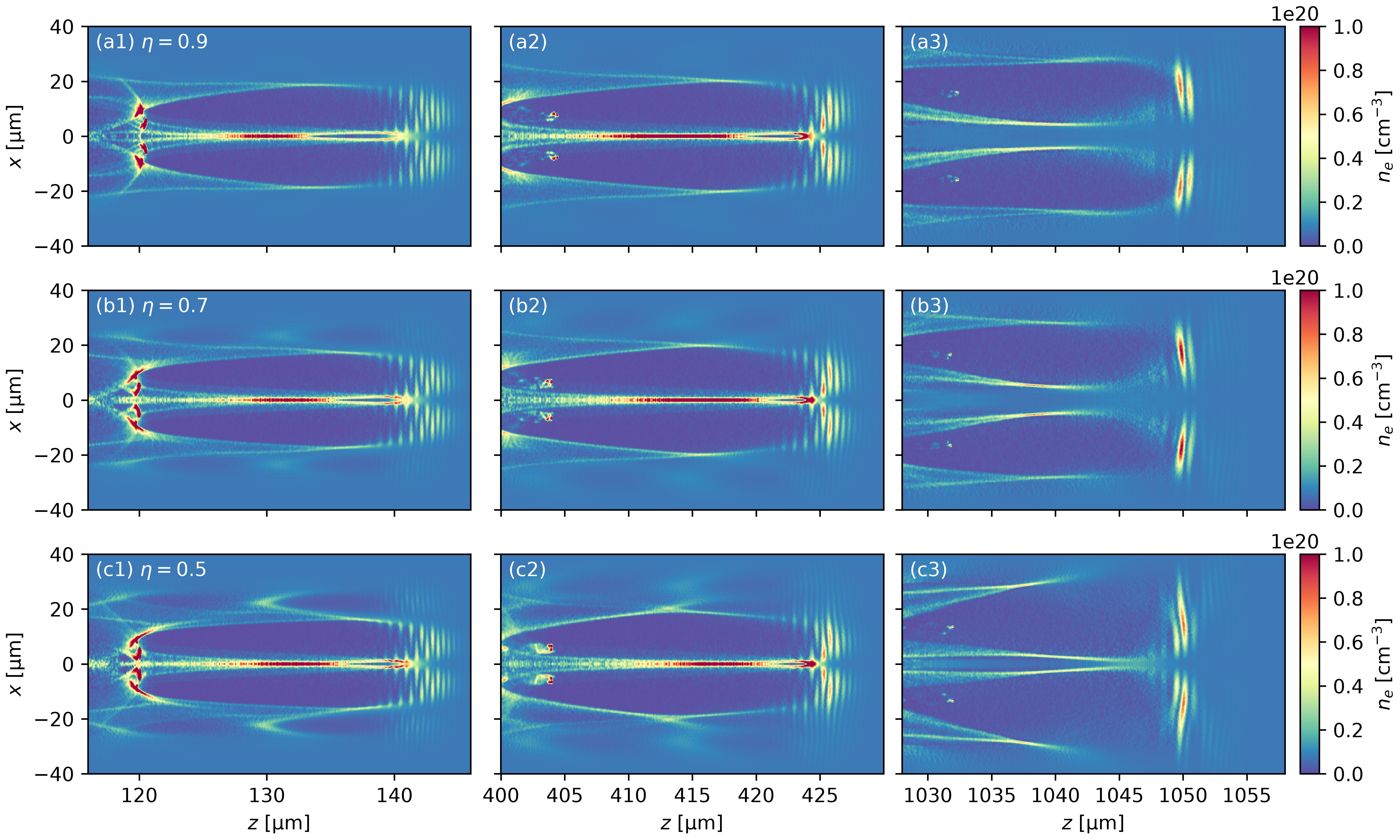}
    \caption{\label{fig:purity_evolution}Time evolution of the wakefield for different degrees of mode purity. Row (a) corresponds to $\eta = 0.9$, row (b) to $\eta = 0.7$ and row (c) to $\eta = 0.5$. Note the existence of additional buckets for lower values of $\eta$ that vanish over the course of the full propagation distance.}
\end{figure*}

In the case of $\eta = 0.9$, the structure of the generated wakefield is rather comparable to the case of a single mode (cf. Fig. \ref{fig:purity_evolution}).
We observe for lower values of $\eta$ that additional structures are formed outside of the main donut cavity. Looking into further time evolution, these vanish over the course of the acceleration length. The final resulting wakefields exhibit the simple donut structure, albeit with some minor modifications at the outer edge.

\begin{figure}
    \centering
    \includegraphics[width=\columnwidth]{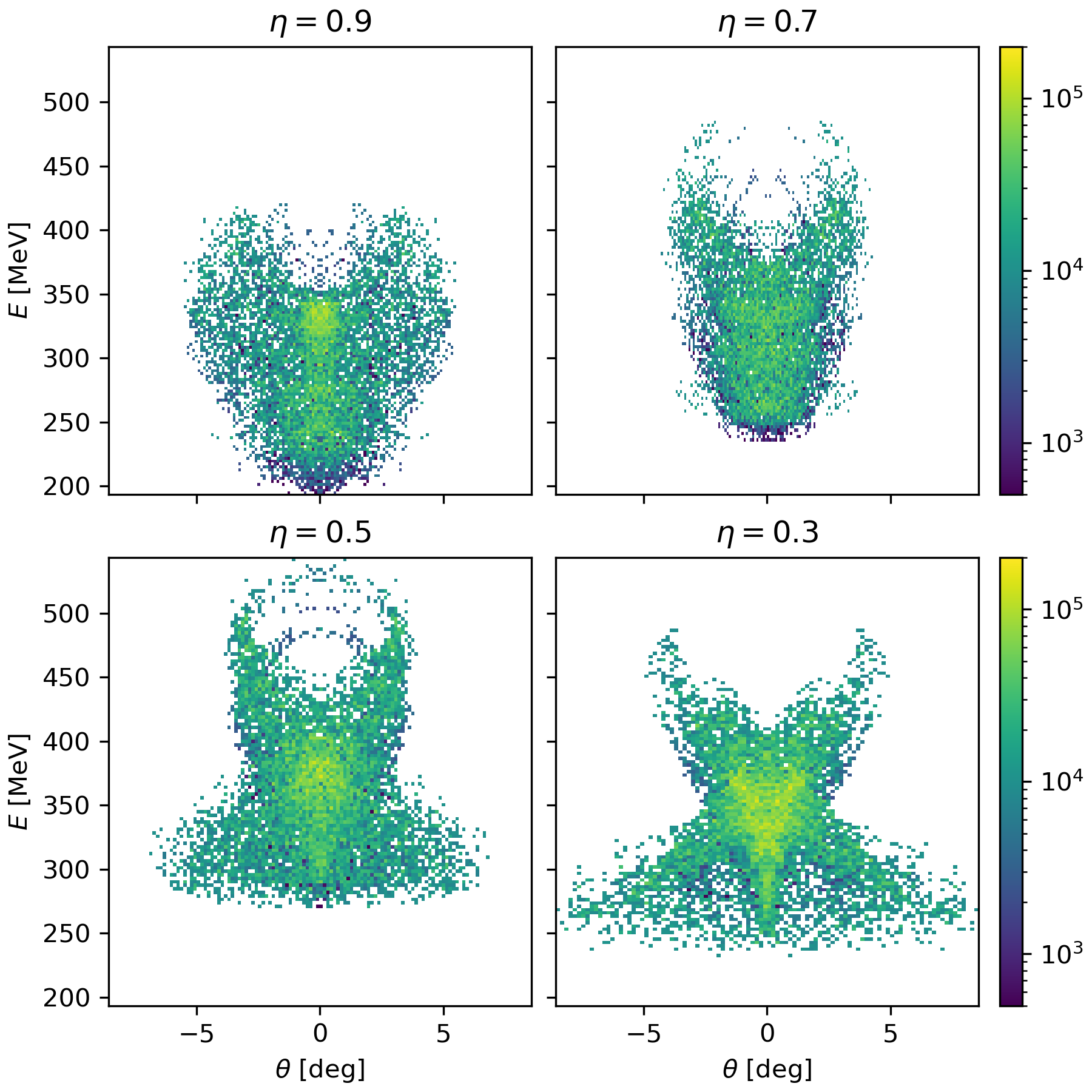}
    \caption{\label{fig:purity} Angle-energy phase space for different mode purity with constant pulse energy. Note that the two modes considered here have no phase difference.}
\end{figure}

From $\eta = 0.9$ to $\eta = 0.5$, an increase in particle energy is observed (cf. Fig. \ref{fig:purity}). For $\eta < 0.5$ the energy decreases and converges towards the single-mode result for (1,1). The apparent increase in energy for certain mixed modes is due to the respective (initial) phase of the two modes: an additional simulation with $\eta = 0.5$, where the phase difference between the two modes is $\pi / 2$ (not shown here), yields maximum energies of approx. 250 MeV instead of the $~ 500$ MeV for in-phase modes. As discussed in more detail in Sec. \ref{sec:ana}, the Gouy phase depends on $\ell$ and $p$, so also modes starting in phase acquire and effective phase difference during propagation.
The simulations demonstrate that wakefield acceleration based on multi-mode LG pulses is complicated especially for cases, where $a_0 \gg 1$.
In particular, the merging of the different lobes will not only depend on the phase between the modes, but also the homogeneity of the plasma target which further determines the propagation of the wakefield structure. Using the merging of multiple beamlets has also been previously proposed to produce higher-intensity fields inside a plasma, e.g. by Chen \textit{et al.} in Ref. \cite{Chen2025}.

\section{Analytical model} \label{sec:ana}

To better understand the effect of mode purity on the intensity profile, we perform some analysis within linear theory. The nonlinear effects of the high-intensity pulses in plasma are more challenging to model and will be subject of a future extended study.
In the notation of Vieira from \cite{Vieira2014}, we can write the laser potential as
\begin{align}
    a_L (\xi, r) = a_0 a_\parallel(\xi) a_r(r) \; ,
\end{align}
where $a_0$ denotes the peak value and $a_\parallel$ and $a_r$ the longitudinal and transverse contributions for each mode ($\xi = z -ct$ being the co-moving coordinate). For our study, we neglect any contributions of $a_\parallel$ since we assume it to be the same for both modes and instead focus on the radial dependence for mode purity.

For a given mode $(\ell, p)$, we write
\begin{align}
    a_r = e^{i \ell \theta} u_{\ell p} \; ,
\end{align}
where
\begin{align}
    u_{\ell p} (r) = \sqrt{\frac{p!}{(p + |\ell|)!}} r^{|\ell|/2} L_{p}^{|\ell|} (r) e^{-r/2}
\end{align}
with the associated Laguerre polynomials $L_{p}^{|\ell|}$.
Considering the mixed mode consisting of (1,0) and (1,1) from the previous section, we may write
\begin{align}
    a_r = e^{i \theta} \left[ \sqrt{\eta} u_{10} (r) + \sqrt{1 - \eta} e^{i \varphi} u_{11}(r) \right] \; .
\end{align}
Here, $\varphi$ denotes the relative phase between the two modes. In the preceding simulations we have considered $\varphi = 0$ for simplicity.
In the case of the two considered modes, the $u_{\ell p}$ terms become
\begin{align}
    u_{10} (r) &= \sqrt{r} e^{-r/2} \; , &&
    u_{11} (r) &= \sqrt{\frac{r}{2}} (2 - r) e^{-r/2} \; .
\end{align}
Inserting these expressions, we obtain
\begin{align}
    |a_r|^2  =& r e^{-r} \left[ \eta + \frac{1}{2} (1 - \eta) (2 - r)^2 \right. \notag \\ 
    & \left. + \sqrt{2 \eta (1 - \eta)} (2 - r) \cos(\varphi) \right] \; .
\end{align}

This gives the radial structure of the modes and their superposition. As stated above, one previously neglected degree of freedom is the phase $\varphi$. Besides the initial respective phase between the two modes, the Gouy phase of the two pulses depends on the mode order \cite{Martinez2015}. A mode of order $(\ell, p)$ acquires a Gouy phase of $(2 p + |\ell| + 1) \arctan[ (z - z_f) / z_R]$, where $z_f$ denotes the focal position and $z_R$ the Rayleigh length. This means that two initially in-phase modes can develop differently over time, modifying the accelerating wakefield structure over the course of its propagation.

\begin{figure}
    \centering
    \includegraphics[width=\columnwidth]{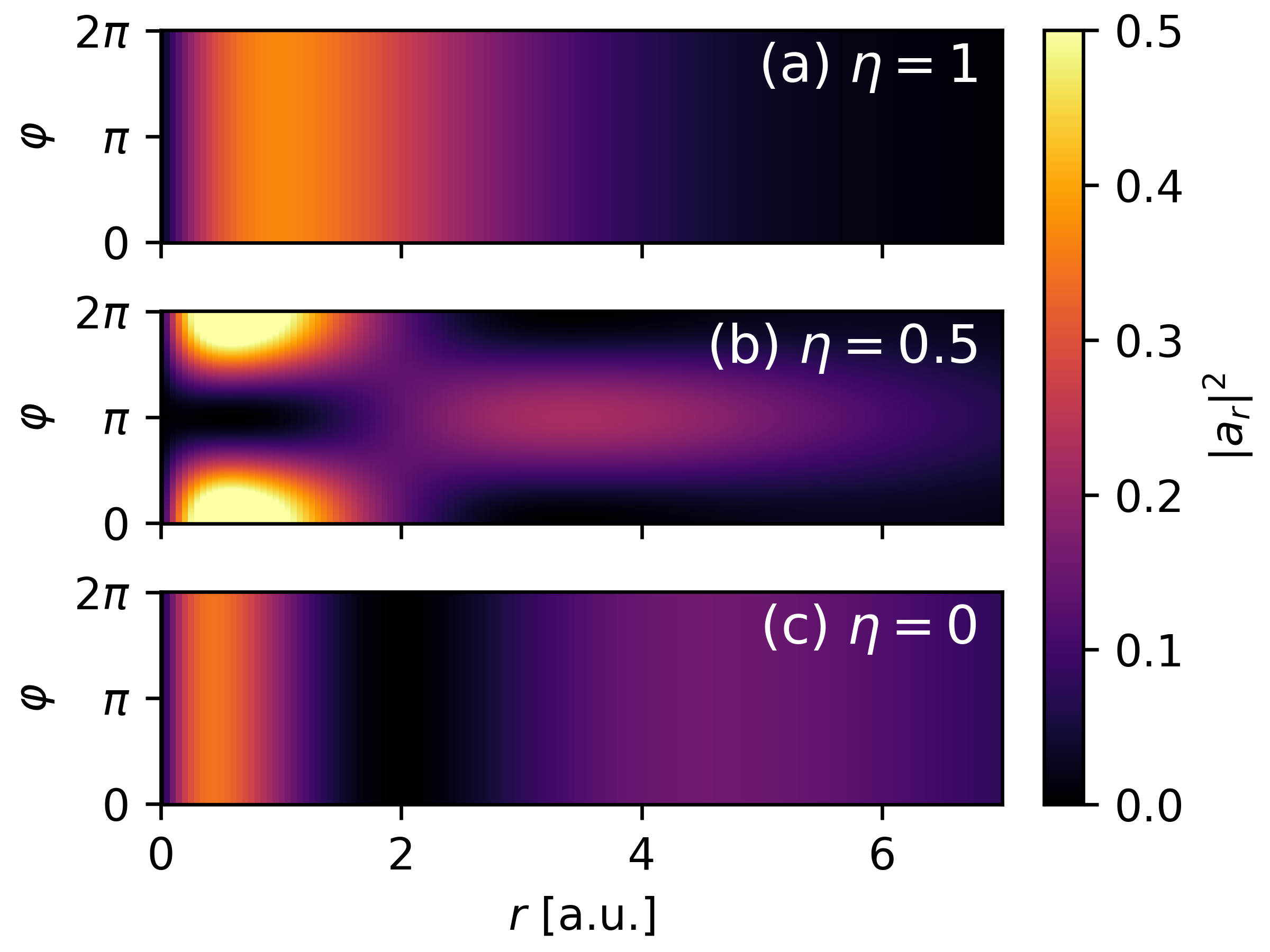}
    \caption{\label{fig:phase_evolution} Radial profiles of $|a_r|^2$ for different mode purities over the course of the phase evolution $\varphi$. Note that the colorbar is clipped for better visibility.}
\end{figure}

We depict the radial profile and the evolution with respect to $\varphi$ for the two pure modes and the mixed mode with $\eta = 0.5$ in Fig. \ref{fig:phase_evolution}.
The maximum of $|a_r|^2$ moves outward for the mixed mode before returning to the initial configuration once $\varphi \to 2\pi$ which would change the equilibrium position of the positron beam being accelerated.
Again, it has to be noted that this result is obtained for linear theory without the presence of plasma. For $a_0 \gg 1$, the simulations indicate (cf. Fig. \ref{fig:purity_evolution}) that mode impurities can be mitigated to some extent, but some differences with respect to pure modes remain.
Nevertheless, the phase difference between the different modes modifies the effective accelerating fields.

\section{Conclusion}

We have studied the influence of mode order and purity of Laguerre-Gaussian laser pulses for positron acceleration using particle-in-cell simulations. The superposition of different modes can be beneficial in terms of final positron energy, but it is very sensitive to the phase difference between the modes.
Accordingly, the results should be taken with care, as an experimentally produced Laguerre-Gaussian pulse will likely have a more complicated mode decomposition. Moreover, the propagation will strongly depend on the strength of the driven wake, as well as the target in terms of general density and its homogeneity. The filamentation and/or merging of the different wakefield lobes can be substantially different for a different choice of parameters. The linear analytical theory also indicates that the propagation of multi-mode pulses is complicated by the fact that different modes acquire different Gouy phases, which in turn affect the wakefield structure. 
A next step will be to extend the analytical theory to stronger pulses and plasma, where nonlinear effects become important.

\begin{acknowledgments}
    This project has received funding from the European Union’s Horizon Europe research and innovation programme under grant agreement no 101096317.  UK participants in the Horizon Europe Project V4F are supported by UKRI grant number 10062154 (MODUS). 
    The author gratefully acknowledges the Gauss Centre for Supercomputing e.V. \cite{GCS} for funding this project (spaf) by providing computing time through the John von Neumann Institute for Computing (NIC) on the GCS Supercomputer JUWELS at J\"ulich Supercomputing Centre (JSC).
    L.R. would like to thank M. Büscher and T.C. Wilson for fruitful discussions.
    Claude Sonnet/Opus 5.5 (Anthropic) have been used for language editing and for cross-checking results. The author reviewed all output and takes full responsibility for the manuscript.
\end{acknowledgments}

\bibliography{positrons_bib}

\end{document}